%% file: main.tex
\documentclass[letterpaper, preprint, comicneue]{aastex63}

\usepackage{lineno}
\usepackage[T1]{fontenc}
\usepackage{fontawesome}
\usepackage{color}
\usepackage{amsmath}
\usepackage{natbib}
\usepackage{ctable}
\usepackage{bm}
\usepackage[normalem]{ulem} 
\usepackage{xspace}
\usepackage{paralist}
\usepackage{fontawesome}
\usepackage{multirow}

\let\oldbibliography\thebibliography 
\renewcommand{\thebibliography}[1]{%
  \oldbibliography{#1}%
  \setlength{\itemsep}{0pt}%
  \setlength{\parsep}{0pt}%
  \setlength{\parskip}{0pt}%
  \setlength{\bibsep}{0ex}
  \raggedright
}
\defcitealias{simbig_letter}{H22a}
\defcitealias{simbig_mock}{H23}

\let\oldAA\AA
\renewcommand{\AA}{\text{\normalfont\oldAA}}

\newcommand{\lya}{Ly$\alpha$}
\newcommand{\lyb}{Ly$\beta$}
\newcommand{\picca}{$\mathtt{picca}$}

\newcommand{\bitem}{\begin{itemize}}
\newcommand{\eitem}{\end{itemize}}
\newcommand{\beq}{\begin{equation}}
\newcommand{\eeq}{\end{equation}}

\newcommand{\lyacolore}{\texttt{Ly$\alpha$CoLoRe}}

\definecolor{orange}{rgb}{1,0.5,0}

\begin{document} \sloppy\sloppypar\frenchspacing 

\title{Reconstructing Quasar Spectra and Measuring the Ly$\alpha$ Forest with {\sc SpenderQ}}

\newcounter{affilcounter}
\author[0000-0003-1197-0902]{ChangHoon Hahn}
\altaffiliation{chhahn@arizona.edu}
\affil{Steward Observatory, University of Arizona, 933 N. Cherry Avenue, Tucson, AZ 85721, USA}
\affil{Department of Astrophysical Sciences, Princeton University, Princeton NJ 08544, USA} 
\affil{Department of Astronomy, The University of Texas at Austin, 2515 Speedway, Stop C1400, Austin, TX 78712, USA}

\author[0000-0003-3142-233X]{Satya Gontcho A Gontcho}
\affiliation{Lawrence Berkeley National Laboratory, 1 Cyclotron Road, Berkeley, CA 94720}
\affiliation{University of Virginia, Department of Astronomy, Charlottesville, VA 22904, USA}

\author{Peter Melchior}
\affil{Department of Astrophysical Sciences, Princeton University, Princeton NJ 08544, USA} 

\author[0000-0002-9136-9609]{Hiram~K.~Herrera-Alcantar}
\affil{IRFU, CEA, Universit\'{e} Paris-Saclay, F-91191 Gif-sur-Yvette, France}
\affil{Institut d'Astrophysique de Paris. 98 bis boulevard Arago. 75014 Paris, France}

\author{Jessica Nicole~Aguilar}
\affil{Lawrence Berkeley National Laboratory, 1 Cyclotron Road, Berkeley, CA 94720, USA}

\author[0000-0001-6098-7247]{Steven~Ahlen} 
\affil{Department of Physics, Boston University, 590 Commonwealth Avenue, Boston, MA 02215 USA}

\author[0000-0001-9712-0006]{Davide~Bianchi}
\affil{Dipartimento di Fisica ``Aldo Pontremoli'', Universit\`a degli Studi di Milano, Via Celoria 16, I-20133 Milano, Italy}
\affil{INAF-Osservatorio Astronomico di Brera, Via Brera 28, 20122 Milano, Italy}

\author{David~Brooks}
\affil{Department of Physics \& Astronomy, University College London, Gower Street, London, WC1E 6BT, UK}

\author{Todd~Claybaugh}
\affil{Lawrence Berkeley National Laboratory, 1 Cyclotron Road, Berkeley, CA 94720, USA}

\author[0000-0002-1769-1640]{Axel~de la Macorra}
\affil{Instituto de F\'{\i}sica, Universidad Nacional Aut\'{o}noma de M\'{e}xico,  Circuito de la Investigaci\'{o}n Cient\'{\i}fica, Ciudad Universitaria, Cd. de M\'{e}xico  C.~P.~04510,  M\'{e}xico}

\author[0000-0002-4928-4003]{Arjun~Dey}
\affil{NSF NOIRLab, 950 N. Cherry Ave., Tucson, AZ 85719, USA}

\author{Peter~Doel}
\affil{Department of Physics \& Astronomy, University College London, Gower Street, London, WC1E 6BT, UK}

\author[0000-0002-2890-3725]{Jaime E.~Forero-Romero}
\affil{Departamento de F\'isica, Universidad de los Andes, Cra. 1 No. 18A-10, Edificio Ip, CP 111711, Bogot\'a, Colombia}
\affil{Observatorio Astron\'omico, Universidad de los Andes, Cra. 1 No. 18A-10, Edificio H, CP 111711 Bogot\'a, Colombia}

\author{Gaston~Gutierrez}
\affil{Fermi National Accelerator Laboratory, PO Box 500, Batavia, IL 60510, USA}

\author[0000-0002-6024-466X]{Mustapha~Ishak}
\affil{Department of Physics, The University of Texas at Dallas, 800 W. Campbell Rd., Richardson, TX 75080, USA}

\author[0000-0002-0000-2394]{Stephanie~Juneau}
\affil{NSF NOIRLab, 950 N. Cherry Ave., Tucson, AZ 85719, USA}

\author[0000-0002-8828-5463]{David~Kirkby}
\affil{Department of Physics and Astronomy, University of California, Irvine, 92697, USA}

\author[0000-0003-3510-7134]{Theodore~Kisner}
\affil{Lawrence Berkeley National Laboratory, 1 Cyclotron Road, Berkeley, CA 94720, USA}

\author[0000-0001-6356-7424]{Anthony~Kremin}
\affil{Lawrence Berkeley National Laboratory, 1 Cyclotron Road, Berkeley, CA 94720, USA}

\author{Andrew~Lambert}
\affil{Lawrence Berkeley National Laboratory, 1 Cyclotron Road, Berkeley, CA 94720, USA}

\author[0000-0003-1838-8528]{Martin~Landriau}
\affil{Lawrence Berkeley National Laboratory, 1 Cyclotron Road, Berkeley, CA 94720, USA}

\author[0000-0001-7178-8868]{Laurent~Le~Guillou}
\affil{Sorbonne Universit\'{e}, CNRS/IN2P3, Laboratoire de Physique Nucl\'{e}aire et de Hautes Energies (LPNHE), FR-75005 Paris, France}

\author[0000-0003-4962-8934]{Marc~Manera}
\affil{Departament de F\'{i}sica, Serra H\'{u}nter, Universitat Aut\`{o}noma de Barcelona, 08193 Bellaterra (Barcelona), Spain}
\affil{Institut de F\'{i}sica d’Altes Energies (IFAE), The Barcelona Institute of Science and Technology, Edifici Cn, Campus UAB, 08193, Bellaterra (Barcelona), Spain}

\author{Ramon~Miquel}
\affil{Instituci\'{o} Catalana de Recerca i Estudis Avan\c{c}ats, Passeig de Llu\'{\i}s Companys, 23, 08010 Barcelona, Spain}
\affil{Institut de F\'{i}sica d’Altes Energies (IFAE), The Barcelona Institute of Science and Technology, Edifici Cn, Campus UAB, 08193, Bellaterra (Barcelona), Spain}

\author[0000-0002-2733-4559]{John~Moustakas}
\affil{Department of Physics and Astronomy, Siena College, 515 Loudon Road, Loudonville, NY 12211, USA}

\author{Adam~D.~Myers}
\affil{Department of Physics \& Astronomy, University  of Wyoming, 1000 E. University, Dept.~3905, Laramie, WY 82071, USA}

\author[0000-0002-1544-8946]{Gustavo~Niz}
\affil{Departamento de F\'{\i}sica, DCI-Campus Le\'{o}n, Universidad de Guanajuato, Loma del Bosque 103, Le\'{o}n, Guanajuato C.~P.~37150, M\'{e}xico}
\affil{Instituto Avanzado de Cosmolog\'{\i}a A.~C., San Marcos 11 - Atenas 202. Magdalena Contreras. Ciudad de M\'{e}xico C.~P.~10720, M\'{e}xico}

\author[0000-0003-3188-784X]{Nathalie~Palanque-Delabrouille}
\affil{IRFU, CEA, Universit\'{e} Paris-Saclay, F-91191 Gif-sur-Yvette, France}
\affil{Lawrence Berkeley National Laboratory, 1 Cyclotron Road, Berkeley, CA 94720, USA}

\author{Claire~Poppett}
\affil{Lawrence Berkeley National Laboratory, 1 Cyclotron Road, Berkeley, CA 94720, USA}
\affil{Space Sciences Laboratory, University of California, Berkeley, 7 Gauss Way, Berkeley, CA  94720, USA}
\affil{University of California, Berkeley, 110 Sproul Hall \#5800 Berkeley, CA 94720, USA}

\author[0000-0001-7145-8674]{Francisco~Prada}
\affil{Instituto de Astrof\'{i}sica de Andaluc\'{i}a (CSIC), Glorieta de la Astronom\'{i}a, s/n, E-18008 Granada, Spain}

\author[0000-0001-6979-0125]{Ignasi~P\'erez-R\`afols}
\affil{Departament de F\'isica, EEBE, Universitat Polit\`ecnica de Catalunya, c/Eduard Maristany 10, 08930 Barcelona, Spain}

\author{Graziano~Rossi}
\affil{Department of Physics and Astronomy, Sejong University, 209 Neungdong-ro, Gwangjin-gu, Seoul 05006, Republic of Korea}

\author[0000-0002-9646-8198]{Eusebio~Sanchez}
\affil{CIEMAT, Avenida Complutense 40, E-28040 Madrid, Spain}

\author{David~Schlegel}
\affil{Lawrence Berkeley National Laboratory, 1 Cyclotron Road, Berkeley, CA 94720, USA}

\author{Michael~Schubnell}
\affil{Department of Physics, University of Michigan, 450 Church Street, Ann Arbor, MI 48109, USA}
\affil{University of Michigan, 500 S. State Street, Ann Arbor, MI 48109, USA}

\author[0000-0002-6588-3508]{Hee-Jong~Seo}
\affil{Department of Physics \& Astronomy, Ohio University, 139 University Terrace, Athens, OH 45701, USA}

\author{David~Sprayberry}
\affil{NSF NOIRLab, 950 N. Cherry Ave., Tucson, AZ 85719, USA}

\author[0000-0003-1704-0781]{Gregory~Tarl\'{e}}
\affil{University of Michigan, 500 S. State Street, Ann Arbor, MI 48109, USA}

\author{Benjamin ~A.~Weaver}
\affil{NSF NOIRLab, 950 N. Cherry Ave., Tucson, AZ 85719, USA}

\author[0000-0002-6684-3997]{Hu~Zou}
\affil{National Astronomical Observatories, Chinese Academy of Sciences, A20 Datun Road, Chaoyang District, Beijing, 100101, P.~R.~China}

\begin{abstract}
    Quasar spectra carry the imprint of foreground intergalactic medium (IGM) 
    through absorption features.
    In particular, absorption caused by neutral hydrogen gas, the
    ``\lya~forest,'' is a key spectroscopic tracer for cosmological analyses 
    used to measure cosmic expansion and test physics beyond the 
    standard model. 
    Despite their importance, current methods for measuring \lya~absorption
    cannot directly derive the intrinsic quasar continuum and make strong 
    assumptions on its shape, thus distorting the measured 
    \lya~clustering.
    We present {\sc SpenderQ}, a ML-based approach for directly reconstructing 
    the intrinsic quasar spectra and measuring the \lya~forest from observations. 
    {\sc SpenderQ} uses the {\sc Spender} spectrum autoencoder to learn a compact and 
    redshift-invariant latent encoding of quasar spectra, combined with an iterative
    procedure to identify and mask absorption regions.
    To demonstrate its performance, we apply {\sc SpenderQ} to 400,000 synthetic 
    quasar spectra created to validate the Dark Energy Spectroscopic 
    Instrument Year 1 \lya~cosmological analyses.
    {\sc SpenderQ} accurately reconstructs the true intrinsic quasar spectra, 
    including the broad \lyb, \lya, SiIV, CIV,  and CIII emission lines.
    Redward of \lya, {\sc SpenderQ} provides percent-level reconstructions of the 
    true quasar spectra.
    Blueward of \lya, {\sc SpenderQ} reconstructs the true spectra to
    < 5\%. 
    {\sc SpenderQ} reproduces the shapes of individual 
    quasar spectra more robustly than the current state-of-the-art. 
    We, thus, expect it will significantly reduce biases in \lya~clustering
    measurements and enable studies of quasars and their physical properties.
    {\sc SpenderQ} also provides informative latent variable encodings that can be used 
    to, e.g., classify quasars with Broad Absorption Lines. 
    Overall, {\sc SpenderQ} provides a new data-driven approach for unbiased 
    \lya~forest measurements in cosmological, quasar, and IGM studies. 
\end{abstract}

\newpage
\input{intro}
\input{methods}

\input{results}
\input{discuss}
\input{summary}

\section*{Acknowledgements}
It's a pleasure to thank 
Daniel Eisenstein, Julien Guy, Stephanie Juneau, and Paul Martini
for useful discussions.
This work was supported by the AI Accelerator program of the Schmidt Futures Foundation. 
This work was substantially performed using the Princeton Research Computing resources
at Princeton University, which is a consortium of groups led by the Princeton Institute 
for Computational Science and Engineering (PICSciE) and Office of Information Technology’s 
Research Computing.

This research used data obtained with the Dark Energy Spectroscopic Instrument (DESI). DESI construction and operations is managed by the Lawrence Berkeley National Laboratory. This material is based upon work supported by the U.S. Department of Energy, Office of Science, Office of High-Energy Physics, under Contract No. DE–AC02–05CH11231, and by the National Energy Research Scientific Computing Center, a DOE Office of Science User Facility under the same contract. Additional support for DESI was provided by the U.S. National Science Foundation (NSF), Division of Astronomical Sciences under Contract No. AST-0950945 to the NSF’s National Optical-Infrared Astronomy Research Laboratory; the Science and Technology Facilities Council of the United Kingdom; the Gordon and Betty Moore Foundation; the Heising-Simons Foundation; the French Alternative Energies and Atomic Energy Commission (CEA); the National Council of Humanities, Science and Technology of Mexico (CONAHCYT); the Ministry of Science and Innovation of Spain (MICINN), and by the DESI Member Institutions: \url{www.desi.lbl.gov/collaborating-institutions}. The DESI collaboration is honored to be permitted to conduct scientific research on I’oligam Du’ag (Kitt Peak), a mountain with particular significance to the Tohono O’odham Nation. Any opinions, findings, and conclusions or recommendations expressed in this material are those of the author(s) and do not necessarily reflect the views of the U.S. National Science Foundation, the U.S. Department of Energy, or any of the listed funding agencies.

\appendix
\input{supp}

\bibliography{spenderq} 
\end{document}

%% file: intro.tex
\section{Introduction} 
Quasars or quasi-stellar objects (QSOs) are luminous active galactic nuclei fueled 
by gravitational accretion onto supermassive black holes at the centers of 
galaxies. 
As the most luminous extragalactic sources in the known universe, quasars 
are one of the main ways to study the cosmic large-scale structure in the 
early universe.
Their spectra serve as cosmic lighthouses that probe the properties of the early 
universe through absorption from objects in their foreground. 
In particular, neutral hydrogen in the Inter-Galactic Medium (IGM) imprints a 
dense collection of absorption lines, blueward of the $\lambda = 1216$\r{A} 
\lya~emission line, that make up the so-called ``\lya~forest''. 


As a tracer of neutral hydrogen, the \lya~forest traces the matter distribution 
and clustering in the Universe. 
On large cosmological scales, this enables us to use the three dimensional 
clustering of the \lya~forest to infer the expansion history of the Universe, 
using the Baryon Acoustic Oscillation feature as a ``standard ruler''~\citep[e.g.,][]{busca2013, slosar2013, Bautista2017, dumasdesbourboux2020}. 
They serve as an essential tracer over the redshift range $2 < z < 3$, 
where we currently do not have other reliable spectroscopic probes. 
\lya~absorption also continuously trace the matter distribution along the 
line of the sight of the background quasar on spatial scales 
as small as the spectral resolution.
They serve as a unique probe of clustering on the smallest scales, 
down to few megaparsecs.
This makes the \lya~forest one of the most promising tracers for 
precisely testing physics beyond the standard model: e.g., 
measuring neutrino mass~\citep[e.g.,][]{font-ribera2014} 
and probing the dark sector~\citep[e.g.,][]{bagherian2024, irsic2024}.
Outside of cosmology, \lya~forests also enable us to probe properties of the
IGM (temperature, density, and ionization state) in the early 
Universe~\citep{haardt2012, madau2015, robertson2015, faucher2020}.

Given their broad cosmological and astrophysical applications, the \lya~forest
has been one of the main tracers observed by recent spectroscopic galaxy surveys:
e.g., the Baryon Oscillation Spectroscopic Survey~\citep[BOSS;][]{eisenstein2011} 
and the extended BOSS~\citep[eBOSS;][]{ahumada2020}. 
The Dark Energy Spectroscopic Instrument~\citep[DESI;][]{desi_snowmass, desicollaboration2016, desicollaboration2016a, desi_kp1}, the first 
Stage-IV galaxy survey, is continuing this effort on the 4m Mayall 
telescope at Kitt Peak National Observatory and with 5000 fiber 
positioners~\citep{guy2023, miller2023, silber2023, schlafly2023, poppett2024}.
DESI has already more than quadruple the number of quasar spectra 
with \lya~from previous surveys, to roughly a million spectra.

To measure the \lya~forest and the full amplitude of its 
absorption, accurate knowledge of the original intrinsic quasar spectra 
(or continuum) is necessary.
Yet, we only observe quasar spectra with \lya~absorption. 
As a result, the quasar continuum must be inferred from the absorbed 
spectra with only pieces of the unabsorbed spectra.
This is only made more difficult by the higher neutral hydrogen fraction 
and density at higher redshifts.
For some quasars, we observe little to none of their intrinsic spectra. 

The current state-of-the-art, \picca\footnote{\url{https://github.com/igmhub/picca/}},
entirely forgoes deriving the intrinsic quasar spectra, $C_q(\lambda)$. 
Instead, it derives the product 
$\overline{F}(\lambda)C_q(\lambda)$, i.e. the intrinsic spectrum multiplied
by the mean transmission fraction, which corresponds to the expected flux 
of the continuum.
It further assumes that the product equates to a universal function in
rest-frame, $G(\lambda/(1+z_q))$, and a polynomial with two 
free parameters $(a_q, b_q)$:
$\overline{F}(\lambda) C_q(\lambda) = G(\lambda/(1+z_q)) (a_q + b_q \log \lambda)$~\citep{dumasdesbourboux2020}.
This assumption does not reflect the range of continuum
shapes in region ``A'' (the \lya~range from $\lambda/(1+z_q) = 1040$ to 1205\r{A}).
\picca~also assumes that the flux transmission 
field,
$\delta_q(\lambda) = f_q(\lambda) / (\overline{F}(\lambda) C_q(\lambda)) -1$, has zero mean and slope.
These assumptions distort the \lya~clustering measurements: 
e.g., they suppress the power spectrum by a factor of two on large 
scales with  $k < 0.1\,h^{-1}{\rm Mpc}$~\citep{blomqvist2015, karacayli2020, karacayli2022, karacayli2024, ravoux2023, debelsunce2024, karim2024}.

Many previous efforts have attempted to go beyond estimating 
$\bar{F}C_q$ and measure the unabsorbed continuum. 
For example, some works tried to exploit the correlation between the 
shape of the continuum on the red and blue side of \lya~to predict 
the continuum using principal component analysis~\citep[e.g.,][]{suzuki2005, paris2011, lee2012, davies2018}. 
More recent works have improved on this approach using ML. 
\cite{liu2021} trained a neural network on Hubble Space 
Telescope/Cosmic Origin Spectrograph (COS) quasar spectra to predict the
continuum in region A. 
\cite{turner2024} used a similar approach with a convolutional neural 
network trained on a combination of COS spectra and DESI Year 5 
mock spectra.
While these supervised learning approaches show promising performance, 
they require training on labeled datasets and are, therefore,
fundamentally limited by ``data mismatch''. 
Any inaccuracies in the observationally derived ground truth or 
limitations in the simulations will degrade or bias their performance.
Alternatively, \cite{sun2023} recently used an unsupervised approach 
with Latent Factor Analysis to learn a model of the continuum. 
The model, however, makes assumptions on the shape of the continuum
and requires modeling the \lya~forest, making it susceptible to 
any limitations in \lya~modeling. 

We present {\sc SpenderQ}, a new method for reconstructing the
intrinsic quasar spectra and measuring the \lya~forest that is fully
data-driven and makes no assumptions on the shape of the continuum.
{\sc SpenderQ} leverages the {\sc Spender} autoencoder
architecture~\citep{melchior2023} to learn compact and redshift-invariant
latent space representations of intrinsic quasar spectra. 
Since the foreground IGM are not correlated with the background quasars,
{\sc Spender} is especially well-suited for being 
insensitive to foregrounds and reconstructing the quasar continuum.
{\sc SpenderQ} further combines {\sc Spender} with an iterative procedure for
identifying and masking \lya~absorption to further improve the fidelity of 
the reconstructions. 
It relaxes the current assumptions on the shape of the quasar continuum to 
produce unbiased \lya~forest measurements. 
Furthermore, as an entirely data-driven and unsupervised approach, 
{\sc SpenderQ} does not require any calibration on simulated data or 
labeled datasets. 
We start by applying {\sc SpenderQ} to synthetic \lya~data created to validate the 
DESI first year \lya~forest BAO analysis.
Then, we compare our quasar reconstructions to the true quasar spectra to demonstrate its performance. 

We begin by describing the {\sc SpenderQ} methodology for quasar spectra reconstruction
(Section~\ref{sec:method}) and a brief description of the synthetic quasar 
spectra (Section~\ref{sec:data}). 
We then present and discuss our results in Section~\ref{sec:results}
and~\ref{sec:discuss}. 

%% file: methods.tex
\begin{figure}
  \centering
  \includegraphics[width=0.8\linewidth]{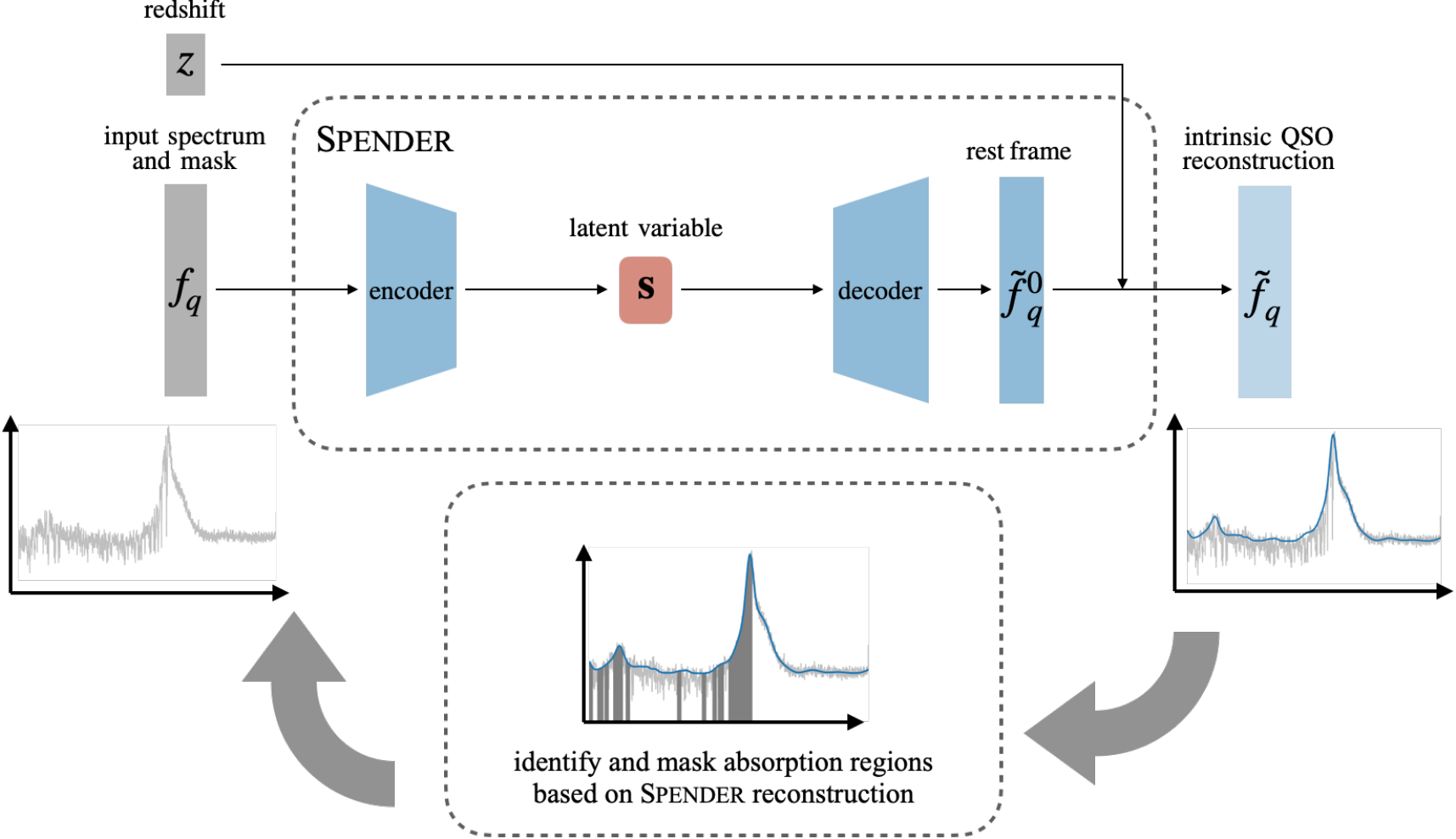}
  \caption{
      {\sc SpenderQ} combines the {\sc Spender} autoencoder with an 
      iterative procedure for masking absorption regions.  
      It uses {\sc Spender} to learn a compact and redshift-invariant 
      latent encoding and reconstruction of the intrinsic quasar 
      spectra from the attenuated spectra and redshift (top).
      Then, it uses the reconstruction to mask absorption regions in the
      spectra for the next iteration (bottom).
      As additional absorption regions are masked with each iteration, 
      {\sc SpenderQ} reconstructions converge to the intrinsic quasar spectra. 
  } \label{fig:arch} 
\end{figure}
\section{\sc SpenderQ} \label{sec:method} 
{\sc SpenderQ} uses the {\sc Spender} autoencoder within an iterative framework
for improving the fidelity of the intrinsic quasar spectrum reconstruction. 
We use {\sc Spender} to learn a compact and redshift-invariant encoding of the
intrinsic quasar spectra from the observed attenuated spectra 
and redshift. 
We use the same architecture as described in \cite{melchior2023} with 
a 10-dimensional latent space\footnote{We find no significant difference in 
performance among models with latent space dimensions of 6 to 12.}
and a rest-frame model that spans the wavelength
$\lambda$ = 800 - 3170 \r{A} with 9,780 spectral elements. 

The foreground IGM does not correlate with the background quasars,
which means that the absorption features in the spectra do not 
correlate with the intrinsic quasar spectra\footnote{
The total overall forest absorption correlates with redshift and, thus, 
the quasar; however, a specific absorption feature does not.}.
In principle, this implies that {\sc Spender} will not learn the absorption
features since the absorptions are effectively random artifacts in the
spectra. 
However, {\sc Spender}, on its own, will learn the average \emph{observed} spectrum, 
including absorption, and therefore underestimate the overall quasar continuum 
because absorption, treated as random noise, is neither additive nor does it 
have a mean of zero.
To account for this effect, we combine {\sc Spender} with an iterative 
procedure for identifying spectral elements with noticeable absorption.

At every iteration, we first train {\sc Spender} using the conventional fidelity
loss as well as similarity and consistency loss, as in \cite{liang2023}. 
The similarity and consistency loss preserves locality in the autoencoding 
process and leads to a redshift-invariant and nondegenerate latent-space 
distribution.
Afterward, we coarsely rebin the quasar spectra, $f_q(\lambda)$, and the 
corresponding {\sc Spender} reconstructions, $\tilde{f}_q(\lambda)$.
The rebinning is done onto a wavelength grid, where the resolution, 
$\Delta \lambda$, is set by the median SNR per pixel of the spectrum 
at $\lambda < 1216$ \r{A} to 
$\Delta \lambda = 8 / {\rm SNR}$ \r{A}, with a minimum $\Delta \lambda > 4$ \r{A}. 
This rebinning is only to identify regions with absorption and 
does not modify the resolution of the training data
or the reconstruction.

Then, we identify spectral elements where $f_q$ lies significantly below 
$\tilde{f}_q$ based on a threshold that depends on $\sigma_q(\lambda)$, the 
observed uncertainty: $f_q < \tilde{f}_q - c \sigma_q$. 
We choose $c$ using a data-driven approach based on the distribution of 
$(\tilde{f}_q - f_q)/\sigma_q$, where we can detect the continuum 
(see Appendix~\ref{app:thresh} for further details).
Over the three wavelength ranges $\lambda < 1026$\r{A}, 
$1026 < \lambda < 1216$\r{A}, and $\lambda > 1216$\r{A} regions,
we use thresholds of $c =0.8, 1.1$, and 3, respectively. 
We mask these spectral elements by setting their weights to zero so 
that they do not contribute to the fidelity loss of the next training epoch.
We repeat the iterations until the reconstructed quasar spectra converge to an 
estimate of the intrinsic spectra $\widetilde{C}_q(\lambda)$, which we find to 
occur after roughly four iterations.  

For the final {\sc SpenderQ} model, we ensemble five independently trained 
{\sc SpenderQ} models. 
Ensembling multiple models improves the robustness of the 
final model~\citep{lakshminarayanan2016, alsing2019, hahn2024}.
Furthermore, we can also use the different models to quantify uncertainties 
on the {\sc SpenderQ} reconstructions. 
We present a schematic illustration of the {\sc SpenderQ} approach and
architecture in Figure~\ref{fig:arch}.

\section{DESI Year 1 Mocks} \label{sec:data} 
In this work, we apply {\sc SpenderQ} to synthetic \lya~data (hereafter; DESI Y1 mocks) 
created to validate the DESI first-year \lya~forest BAO analysis~\citep{DESI_KP6, desi2024_bao}. 
The data are from a realization of the \lyacolore~mocks presented in \cite{Cuceu:2024Validation}. 
First, the \texttt{CoLoRe} package\footnote{\url{https://github.com/damonge/CoLoRe}}\citep{Ramirez_Perez_2022}
is used to generate a set of transmitted flux fraction skewers and positions of quasars associated to them,
based on log-normal Gaussian random fields. 
Next, the skewers are post-processed using the \lyacolore~package\footnote{\url{https://github.com/igmhub/LyaCoLoRe}}~\citep{Farr_2020}
to introduce small-scale corrections and redshift-space distortions. 
The \lyacolore\ package also generates a catalog of high column density systems (HCDs) correlated 
with the \lya\ density field.
Lastly, realistic DESI-like mock quasar spectra are constructed from the \lyacolore~package 
outputs using the \texttt{quickquasars} 
script\footnote{\url{https://github.com/desihub/desisim/blob/main/py/desisim/scripts/quickquasars.py}}~\citep{Herrera-Alcantar:2023its} in the 
\texttt{desisim} package\footnote{\url{https://github.com/desihub/desisim}}.

A subset of mock quasars are randomly selected to closely match the characteristics of the DESI
Year 1 \lya\ sample in terms of: footprint, redshift-magnitude relation, and 
mean SNR per pixel over $\lambda/(1+z_q) = $ 1040 - 1205\r{A}. 
Then, absorption features from astrophysical effects are incorporated into the transmitted flux
fractions of the selected quasars. 
The mocks include astrophysical effects from HCDs, which are modeled 
using a Voigt profile.
They also include Broad Absorption Lines (BALs), randomly introduced into 16\% of the spectra based 
on a set of templates~\citep{Martini:2024upk}, and absorption features from other transition lines
(metals), which are added by re-scaling the \lya\ transmitted flux fraction 
according to each transition and shifting the absorption features to the appropriate wavelengths.
For additional details, we refer readers to Section 2.3 of \cite{herrera2025}.

The transmitted flux fractions with all these features are then multiplied to 
unabsorbed quasar spectrum, which are templates assigned to the quasars based on 
$r$-band magnitude and redshift using the \texttt{SIMQSO} library\footnote{\url{https://github.com/imcgreer/simqso}}~\citep{simqso:2021}.
Instrumental noise is then introduced into the quasar spectra using 
\texttt{specsim}\footnote{\url{https://github.com/desihub/specsim}}~\citep{Kirkby:2016}, which 
emulates the DESI spectrograph and the nominal DESI dark-time program observational conditions.
We refer the reader to \cite{Ramirez_Perez_2022,Farr_2020, Herrera-Alcantar:2023its} and 
\cite{Cuceu:2024Validation} for further details on \texttt{CoLoRe}, \lyacolore, \texttt{quickquasars}, 
and the DESI Y1 mocks, respectively.

For this work, we limit our analysis to quasar spectra with redshifts between 
the range $2.1 < z < 3.5$.
The lower bound corresponds to the redshift limit of the DESI \lya~sample 
in \cite{desi_lya_bao} and ensures that all of the quasars have some rest-frame
wavelength overlap. 
The upper bound is the nominal limit for the DESI \lya~quasar tracers.
Before applying {\sc SpenderQ}, we normalize the spectra to reduce their dynamical 
range. 
We divide them by the median spectral flux within 
$\lambda/(1+z_q)$ = 1600 - 1800\r{A}. 
This range roughly corresponds to the wavelengths  between the CIV and CIII 
emission lines, where there is little absorption overall from CIII. 
In total, our sample consists of 401,820 quasar spectra.

%% file: results.tex
\begin{figure}
  \centering
  \includegraphics[width=\linewidth]{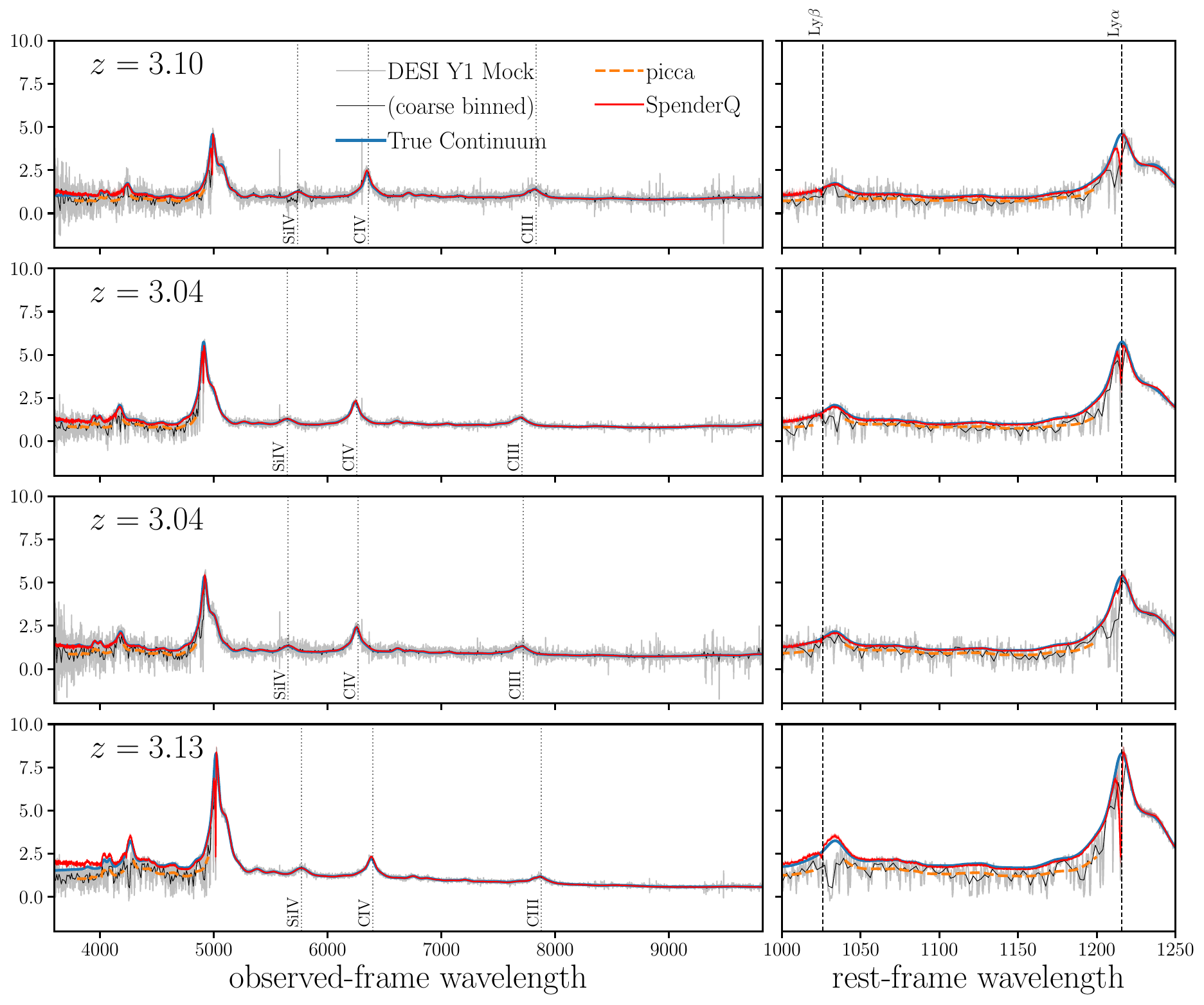}
  \caption{
      {\sc SpenderQ} reconstructed intrinsic quasar spectra (red) compared to
      the DESI Y1 mock spectra (gray). 
      We plot randomly selected high redshift, $z > 3$, quasars with higher
      signal-to-noise to showcase the full \lya~region and for clarity. 
      The left panels present the full spectra over the observed-frame wavelength. 
      The right panels focuses on region A in rest-frame wavelength. 
      We include the true quasar spectra (blue), the coarsely rebinned spectra (black), 
      and the \picca~$\overline{F}(\lambda)C_q(\lambda)$ estimates 
      (orange dashed), for reference. 
      {\sc SpenderQ} reconstructions are in excellent agreement with the true 
      quasar spectra and, unlike \picca, directly estimate the intrinsic quasar 
      spectra, $C_q(\lambda)$.
  } \label{fig:recon}
\end{figure}

\section{Results} \label{sec:results}
We present a comparison of the reconstructed intrinsic quasar spectra 
derived using {\sc SpenderQ} (red) to the DESI Y1 mock spectra (gray)
in Figure~\ref{fig:recon}. 
The spectra and reconstructions are plotted over the full observed-frame 
wavelength range in the left panels while the right panels focus on 
region A in the rest-frame ($\lambda/(1+z_q) =$ 1040 - 1205\r{A}).
All spectra are normalized (Section~\ref{sec:data}). 
We include the true intrinsic quasar continua (blue) and the 
spectra coarsely rebinned to 10\r{A} resolution (black), for comparison. 
For reference, we include the $\mathtt{picca}$ estimates of
$\bar{F}(\lambda)C_q(\lambda)$ over regions A and ``B'' 
($\lambda/(1+z_q)=$920 - 1020\r{A}) in orange dashed. 
We also include the uncertainties of the {\sc SpenderQ} reconstructions
by plotting the range of the predictions from all of the {\sc SpenderQ} models
in the ensemble (shaded red). 
The quasars are randomly selected $z > 3$ quasars with higher signal-to-noise
to showcase the full \lya~region and for clarity.

Overall, {\sc SpenderQ} accurately reconstructs the quasar spectra over the 
full observed wavelength range. 
This is especially clear redward of region A, $\lambda \gtrsim 1200$\r{A}, 
where absorption features are rarer. 
The broad emission lines of \lyb, \lya, SiIV, CIV, and CIII are well reconstructed
and in excellent agreement with the true spectra in all of the selected quasars, 
even the asymmetric profiles of \lya~and \lyb.
In region A, the {\sc SpenderQ} reconstructions successfully ignore 
absorption features and are in excellent agreement with the true quasar continua. 
We emphasize that {\sc SpenderQ} reconstructs the true quasar continua, $C_q$, 
unlike \picca, which can only estimate $\overline{F}C_q$. 

\begin{figure}
  \centering
  \includegraphics[width=\linewidth]{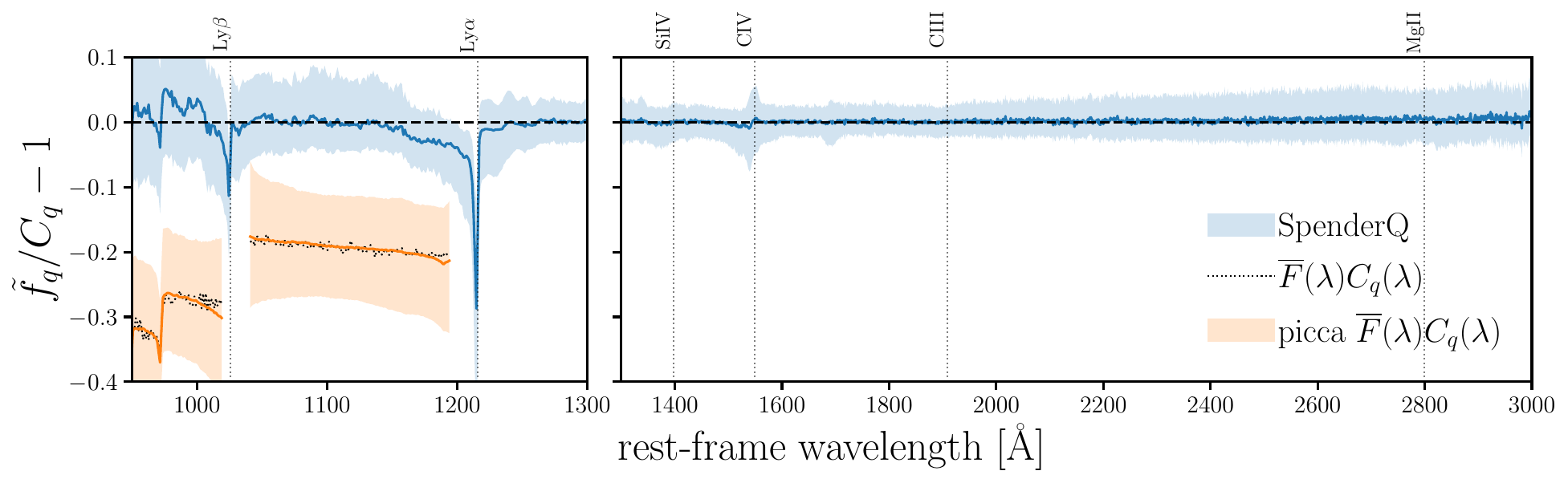}
  \caption{Fractional residual of the {\sc SpenderQ} reconstruction versus 
  the true quasar spectra over a broad rest-frame wavelength range.
  The shaded region represents the 16 and 84$^{\rm th}$ percentiles of the residuals. 
  For reference, we include the residuals of the $\mathtt{picca}$ 
  $\overline{F}(\lambda)C_q(\lambda)$ estimates (orange) and the true $\overline{F}(\lambda)C_q(\lambda)$ (black dotted).
  Redward of \lya, {\sc SpenderQ} provides percent level reconstructions of the 
  true spectra (right). 
  Blueward of \lya, {\sc SpenderQ} reproduces the true spectra 
  with a mean residual of <5\% (left).
  } \label{fig:resid}
\end{figure}

In Figure~\ref{fig:resid}, we present a quantitative assessment of {\sc SpenderQ}
with the fractional residual of the reconstructions versus the true quasar 
spectra, $\tilde{f_q} / C_q - 1$ (blue), evaluated over all the 
mock spectra. 
The shaded region represents the 16 and 84$^{\rm th}$ percentiles of the residuals. 
For reference, we mark the emission line wavelengths in black dashed. 
We also include the residuals of the \picca~
$\overline{F}(\lambda)C_q(\lambda)$ estimates (orange) and the true 
$\overline{F}(\lambda)C_q(\lambda)$ (black dotted), for reference.

Redward of \lya, $\lambda/(1+z_q) \gtrsim$ 1250\r{A}, {\sc SpenderQ} reproduces the true 
spectra to percent level (right panel). 
We note that the performance of {\sc SpenderQ} is worse at longer 
wavelengths, $\lambda/(1+z_q)$ > 2800\r{A}, with a slightly broader distribution 
of residuals.
This is due to the fact that we normalize the spectra by the median flux 
within $\lambda/(1+z_q)=$1600 - 1800\r{A} (Section~\ref{sec:data}). 
Nevertheless, the residuals demonstrate that even without any additional 
postprocessing {\sc SpenderQ} is capable of accurately detecting
and measuring SiV, CIV, CIII, and MgII metal absorption line that 
probe the gas content in the high redshift Universe. 

\begin{figure}
  \centering
  \includegraphics[width=0.7\linewidth]{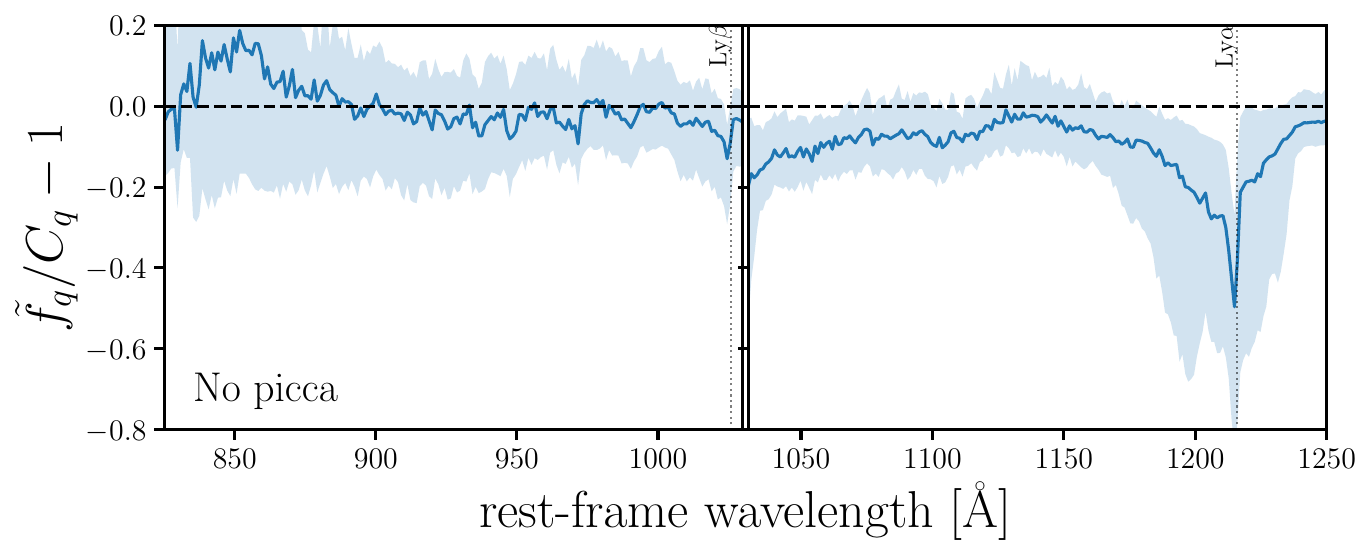}
  \caption{Fractional residual of the {\sc SpenderQ} reconstruction versus 
  the true spectra for quasars with no \picca~reconstruction. 
  \picca~fails to fit $\sim$5\% of the spectra for region A and 
  $\sim$70\% of the spectra for region B where the are fewer than 150 spectral elements.
  The shaded region represent the 16 and 84$^{\rm th}$ percentiles of the residuals. 
  Even for spectra with no \picca~reconstructions, {\sc SpenderQ} 
  produces reconstructs the quasar spectra with $\lesssim$10\% residuals. 
  } \label{fig:nopicca}
\end{figure}

In region A, {\sc SpenderQ} reproduces the true spectra with mean 
residual <5\%. 
In fact, over $\lambda/(1+z_q) =$ 1040 - 1150\r{A}, 
the mean {\sc SpenderQ} residual is <1\%, with no noticeable wavelength dependence. 
Although we refrain from direct comparison of the residuals with \picca, 
as it estimates $\overline{F}C_q$ and not $C_q$, we note that the 
{\sc SpenderQ} reconstructions are more precise.
The shaded region is 1.5$\times$ narrower than the shaded region of 
\picca. 

In region B, blueward of \lyb, $\lambda/(1+z_q) \lesssim$ 1050\r{A}, 
{\sc SpenderQ} performs  noticeably worse. 
This region is affected by both \lya, \lyb, and other Lyman series
absorption and is only available in spectra of quasars with $z > 2.4$. 
Even for the $z > 2.4$ quasars, region B is at the edge of the DESI
spectral range, with lower throughput, and thus is 
noisier~\citep[e.g.,][]{chaussidon2023}.
Nevertheless, {\sc SpenderQ} reconstructions have mean residuals 
of <5\% with similar precision as \picca. 

We note that {\sc SpenderQ}'s performance is despite the fact that we 
do not preprocess the spectra to mask BAL and damped \lya~(DLA)
systems\footnote{This preprocessing is included in \picca.}. 
However, because we do not mask BALs there is a significant residual
in the emission line regions of \lya~and \lyb. 
On the blue side of the lines, there is a $\sim$3\% suppression in the
reconstructions. 
This is caused by the fact that absorption from BAL features can 
correlate with the intrinsic quasars and, thus, is learned by 
{\sc SpenderQ}. 
We discuss this further in Section~\ref{sec:discuss}. 

In addition to directly estimating quasar continua, {\sc SpenderQ} also 
provides reconstructions for all spectra. 
\picca~does not fit a spectrum if it has less than 150 spectral elements 
in the wavelength range. 
This excludes $\sim$5\% of the spectra for region A and 
$\sim$70\% of the spectra for region B.
However, {\sc SpenderQ} can reconstruct the quasar spectra for all quasars.  
In Figure~\ref{fig:nopicca}, we present the {\sc SpenderQ} residuals 
for quasar spectra with no \picca~fits for region B (left) and no fits for region A (right). 
Despite the limited data, {\sc SpenderQ} produces reconstructions with 
residuals of $\sim$10\%. 
{\sc SpenderQ} is able to do this because it leverages many correlated 
features throughout the entire spectra to produce accurate reconstructions 
(see also discussion in \citealt{melchior2023}).

\begin{figure}
  \centering
  \includegraphics[width=\linewidth]{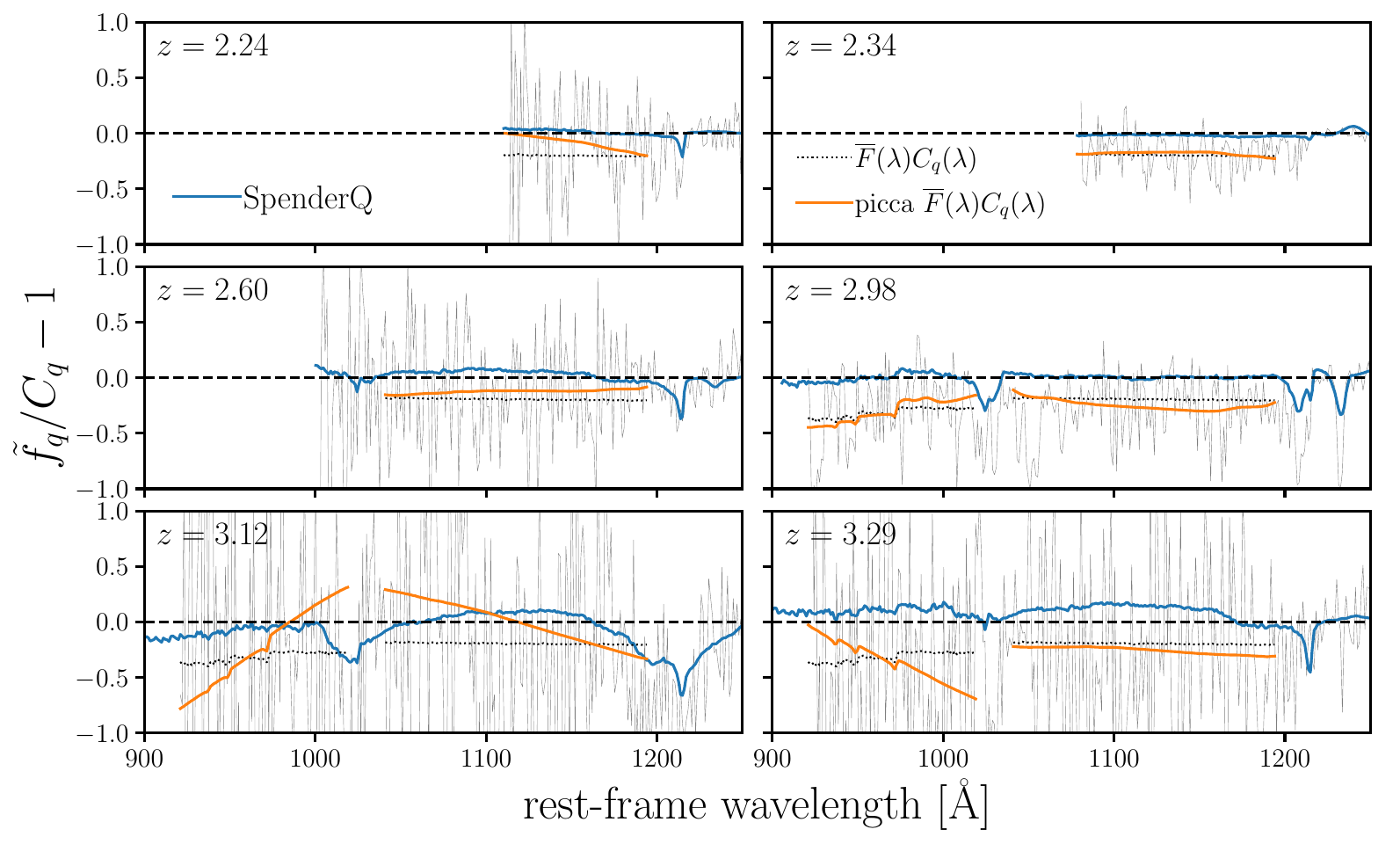}
  \caption{Fractional residual of the {\sc SpenderQ} reconstruction (blue) for 
  individual spectra for random quasars across $2.1 < z < 3.5$.
  We include residuals of the \picca~$\overline{F}C_q$ estimates (orange), 
  the true $\overline{F}C_q$, and the mock observed spectrum (black) 
  for comparison. 
  Even for a small sample of randomly selected quasars, the comparison reveals
  cases where \picca~dramatically misestimates the shape of $\overline{F}C_q$.
  Meanwhile, {\sc SpenderQ} provides more robust reconstructions that 
  accurately capture the shape of the continuum. 
  } \label{fig:recon_indiv}
\end{figure}

In Figure~\ref{fig:recon_indiv}, we further examine {\sc SpenderQ}'s performance; 
this time for individual spectra.
We present the fractional residuals of {\sc SpenderQ} reconstructions for a 
set of spectra randomly sampled across our redshift range (blue). 
For comparison, we include the residuals of the \picca~$\overline{F}C_q$ 
estimates (orange), the true $\overline{F}C_q$ (black dotted), 
and the observed spectrum (black). 
For some of the spectra the \picca~estimates have significantly biased slopes
compared to the true $\overline{F}C_q$, blueward of \lya.
This is particularly the case when region A or B falls near the edge 
of the DESI wavelength range with low signal-to-noise. 
Meanwhile, {\sc SpenderQ} accurately reconstructs the shape of the continua, 
even in the low signal-to-noise regimes.  

\begin{figure}
  \centering
  \includegraphics[width=0.5\linewidth]{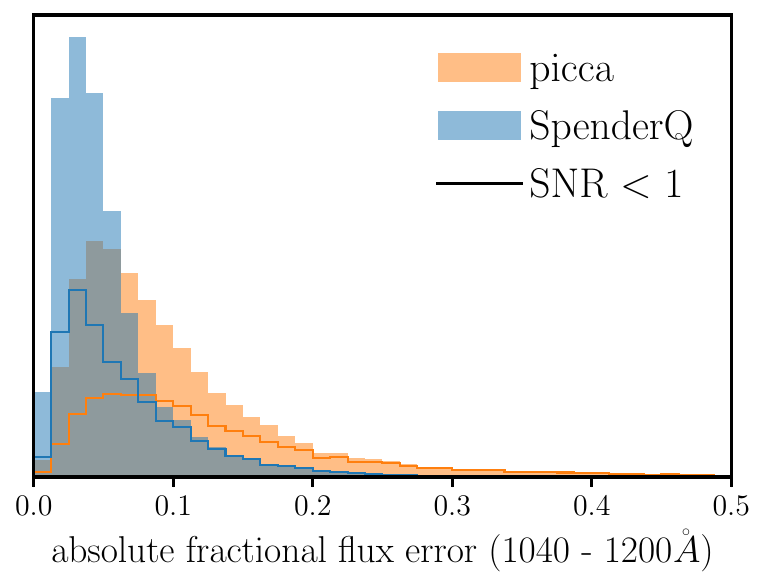}
  \caption{The distribution of absolute fractional flux error (AFFE) 
  of the {\sc SpenderQ} (blue) and the \picca~reconstructions (orange). 
  We calculate the AFFE for all mock spectra with \picca~reconstructions 
  in region A and over the wavelength  range: 1040 - 1200\r{A}. 
  For \picca, we calculate AFFE with respect to $\overline{F}C_q$ as it 
  estimates $\overline{F}C_q$, not $C_q$. 
  We also include the AFFE distributions for spectra with low SNR < 1
  (solid line). 
  The comparison of the AFFE distributions illustrates that 
  {\sc SpenderQ} provides more accurate and precise quasar continuum reconstructions. 
  } \label{fig:affe}
\end{figure}

To quantitatively access {\sc SpenderQ}'s performance on individual 
spectra, we calculate the absolute fractional flux 
error~\citep[AFFE; e.g.,][]{liu2021, turner2024}
of its reconstructions:
\begin{equation}
|\delta C_q| = \frac{\int\limits_{\lambda_1}^{\lambda_2}|\frac{\tilde{f}_q(\lambda) - C_q(\lambda)}{C_q(\lambda)}|\,{\rm d}\lambda}{\int\limits_{\lambda_1}^{\lambda_2} {\rm d}\lambda}. 
\end{equation}
We evaluate the AFFE  over the \lya~region ($\lambda_1 = 1040$ and $\lambda_2 = 1200$\r{A}) for all mock spectra with \picca~reconstructions. 
In Figure~\ref{fig:affe}, we present the distribution of the AFFE for 
{\sc SpenderQ} reconstructions (blue). 
We also present the AFFE distribution for \picca~(orange), where  
we evaluate AFFE for $\bar{F}C_q$ rather than $C_q$ for an 
apples-to-apples comparison. 
Overall, {\sc SpenderQ} reconstructions have significantly lower 
AFFE than \picca~with median AFFE of 0.04 versus 0.08. 
Furthermore, with {\sc SpenderQ} we have fewer reconstructions with 
large AFFE: e.g., less than 1.5\% have AFFE > 0.2 compared to 
$\sim$10\% for \picca. 
We also include in Figure~\ref{fig:affe} the {\sc SpenderQ} 
and \picca~AFFE distributions for spectra with lower SNR (solid lines). 
These spectra have median SNR per pixel < 1 and account for half 
of the spectra. 
The AFFE is higher for lower SNR spectra, for both 
{\sc SpenderQ} and \picca.
However, {\sc SpenderQ} has significantly lower AFFE even for lower 
SNR spectra.
The comparisons in Figure~\ref{fig:recon_indiv} and \ref{fig:affe}
demonstrate:
{\sc SpenderQ} reconstructions are not only accurate and more precise overall, but also more 
robust for individual spectra. 

%% file: discuss.tex
\section{Discussion} \label{sec:discuss}
We demonstrate the advantages of {\sc SpenderQ} over the current
state-of-the-art. 
{\sc SpenderQ} directly estimates the quasar continua, not $\overline{F}(\lambda)C_q(\lambda)$.
Its reconstructions are overall unbiased and more precise.  
It also more accurately reconstructs individual quasar spectra and can be
applied to all spectra, even ones with limited coverage of region A or B. 

Residuals in the quasar continuum bias flux transmission field derived from them. 
This in turn significantly distorts \lya~clustering measurements~\citep{busca2013, slosar2013, delubac2015, blomqvist2015, Bautista2017}.
For example, the amplitude and wavelength dependence of the residual suppress
the monopole of power spectrum by a factor of two on large scales 
($k < 0.1\,h^{-1}{\rm Mpc}$) and also distorts of the shape of all the 
multipoles~\citep{debelsunce2024}. 
With {\sc SpenderQ}, we expect to significantly reduce this distortion and 
reduce the systematics in \lya~clustering analyses. 

Despite the improvements, {\sc SpenderQ} does not accurately reconstruct every 
spectra.
For 0.25\% of the spectra, the median {\sc SpenderQ} fractional residual in
the \lya~range exceed 25\%.
These spectra fall broadly under two categories.
The first are spectra with low signal-to-noise.
99\% of the spectra with higher residuals have SNR < 1.  
The second are spectra with higher SNR but with major broad absorption 
features in the \lya~emission from BALs. 
In the standard \lya~cosmological pipeline, spectra with BAL and DLAs are
identified, then masked or removed by preprocessing steps before any 
reconstruction\footnote{
We include this preprocessing for the \picca~reconstructions in 
Figure~\ref{fig:resid}.}. 
In this work, we opt to keep these contaminants to fully test the capabilities of {\sc SpenderQ}.
BALs come from high-velocity outflows~\citep{wymann1991} 
and correlate with the intrinsic properties of quasar. 
{\sc SpenderQ}, therefore, learns these correlated features from 
these spectra, which impact their reconstruction.
We show a few examples of these spectra in Appendix~\ref{app:limit},  
Figure~\ref{fig:bad_recon}.

\begin{figure}
  \centering
  \includegraphics[width=\linewidth]{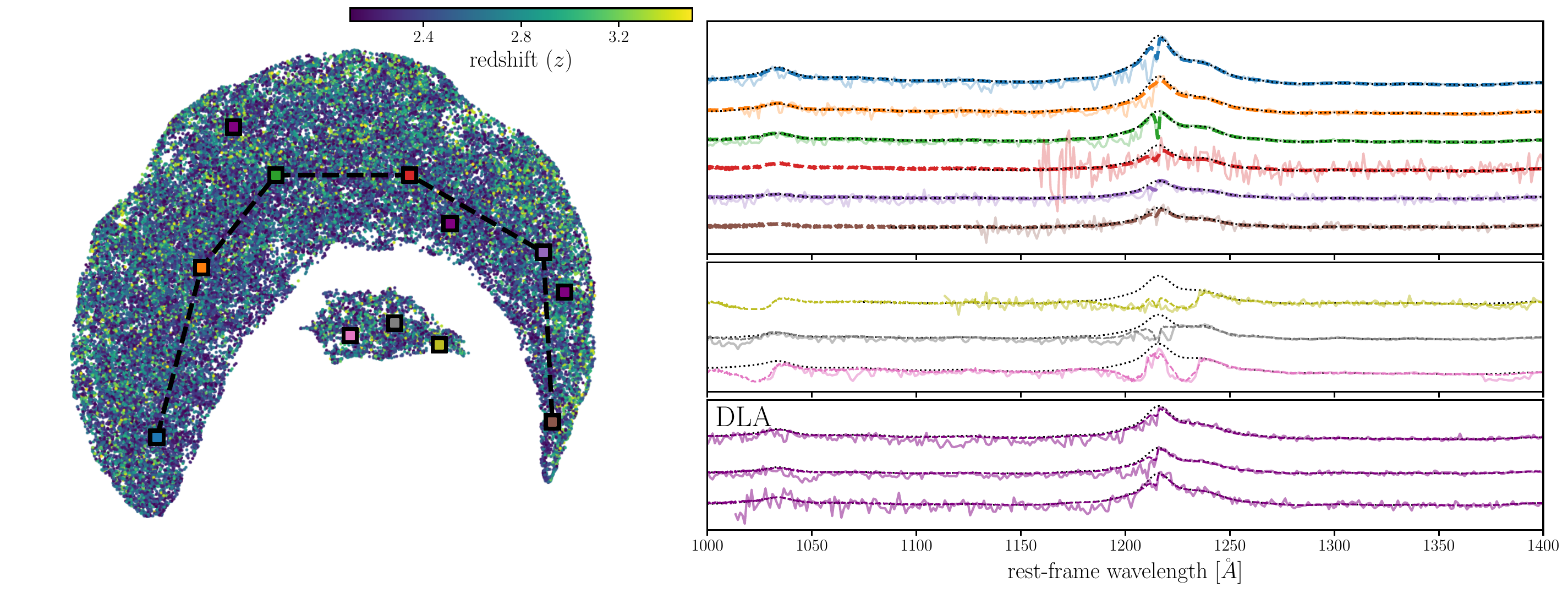}
  \caption{
  {\em Left}: The two-dimensional UMAP embedding of the {\sc SpenderQ} 
  latent variables.
  We mark the redshifts of the quasars through the color bar. 
  The embedding is split into two distinct groups. 
  We find no significant redshift dependence in the embedding, which indicate
  that {\sc SpenderQ} learns a redshift invariant encoding of quasar spectra.
  {\em Right}: We examine the spectra of quasars marked in the UMAP embedding. 
  The top and center panels present spectra from the upper and lower 
  groups, respectively.
  The bottom panel presents spectra with DLAs.
  For reference, we include the true quasar spectra (black dotted) and 
  the {\sc SpenderQ} reconstruction (dashed). 
  The spectra are offset for clarity.
  The spectra in the upper panel show how the overall shape of quasar changes
  along the upper embedding.  
  Meanwhile, the spectra in the lower embedding clearly show broad 
  absorption features. 
  } \label{fig:latents}
\end{figure}

Fortunately for the spectra with BAL features, the latent 
variable space of {\sc SpenderQ} provides useful insights. 
In the left panel of Figure~\ref{fig:latents}, we present the two-dimensional Uniform Manifold Approximation and 
Projection~\citep[UMAP;][]{mcinnes2018} embedding of the {\sc SpenderQ} latent variables.
We color the quasars by their redshifts. 
The embedding reveals a clear separation of the quasars into two distinct 
clusters. 
In the right panels of Figure~\ref{fig:latents}, we present the spectra for 
quasars marked in the UMAP embedding in the upper (top) and lower (center) clusters.
We also present randomly selected spectra with DLAs (bottom).
We offset the spectra for clarity and include the {\sc SpenderQ} 
reconstructions (dashed) and the true continuum (black dotted) for reference. 

Each of the spectra in lower cluster display clear broad 
absorption features in the \lya~emission region. 
Indeed, we confirm that 99\% of the spectra in this cluster have flagged BALs. 
This means we can use the {\sc SpenderQ} latent space to identify spectra 
with BALs and postprocess them to further improve their reconstruction.
Alternatively, we can also apply {\sc SpenderQ} to the preprocessed 
dataset where spectra with BALs are masked or removed. 
This will only improve the performance of {\sc SpenderQ}. 
Meanwhile, the lower right panel illustrates that {\sc SpenderQ} 
reconstructions are robust to DLAs. 
Unlike BALs, DLAs do not correlate with the intrinsic quasar spectra
so {\sc SpenderQ} reconstructions are not impacted by them.

The UMAP reveals additional insights into {\sc SpenderQ}. 
First, the color coding by redshift reveals that there is no significant 
redshift dependence throughout the latent space and confirms that 
{\sc SpenderQ} learns a redshift invariant encoding of the spectra. 
This further explains how {\sc SpenderQ} produces better reconstructions 
in region B where \picca~struggles.
It leverages higher redshift quasar spectra with better coverage
for spectra with lower redshift quasars with little region B coverage. 
To further examine the latent space, we present the spectra of quasars
along the marked track in the upper cluster in the top right panel 
of Figure~\ref{fig:latents}.  
Quasars along this track have intrinsically different spectra with varying 
\lya~emission. 
This demonstrates that structure in the {\sc SpenderQ} latent space is
physically informative and represent intrinsically different quasar  spectra. 
Also, the {\sc SpenderQ} latent space can be used for outlier or anomaly detection 
as in \cite{liang2023, liang2023a, boehm2023}. 

%% file: summary.tex
\section{Summary} \label{sec:summary}
We present {\sc SpenderQ}, a ML-based and fully data-driven method for 
reconstructing the instrinsic quasar spectra and measuring the \lya~forest.
{\sc SpenderQ} leverages the {\sc Spender} autoencoder to learn compact and
redshift-invariant latent space representations of intrinsic quasar spectra.
It combines this with an iterative procedure for identifying and masking 
absorption features to further improve the fidelity of the reconstructions. 
It relaxes any assumptions on the shape of the quasar continuum to produce 
unbiased \lya~forest measurements.

We apply {\sc SpenderQ} to synthetic quasar spectra used to validate the DESI 
first year \lya~forest BAO analysis and demonstrate its  performance:  
\begin{enumerate}
    \item {\sc SpenderQ} accurately reconstructs the true intrinsic quasar 
    spectra --- not $\overline{F}C_q$--- over the full wavelength range, including the broad emission lines (\lyb, \lya, SiIV, CIV,  and CIII).
    For rest-frame $\lambda$ > 1215\r{A}, {\sc SpenderQ} reconstructs 
    the true continuum to percent level. 
    \item Over the \lyb~and \lya~ranges, the {\sc SpenderQ} reconstructions
    overall have <5\% mean residuals compared to the true continuum. 
    For region A, {\sc SpenderQ} is 1.5$\times$ more precise than 
    \picca, the current state-of-the-art. 
    This is despite not preprocessing the spectra to identify and mask BALs and 
    DLAs, as done in standard \lya~clustering analyses.
    \item Beyond its overall performance, {\sc SpenderQ} robustly reconstructs 
    the shape of individual quasar spectra better than \picca, with overall half the AFFE.
    It also reconstructs the continuum of all quasar spectra, even ones with 
    limited data in regions A and B. 
    With these improvements, {\sc SpenderQ} will significantly 
    reduce the biases and distortions in \lya~clustering measurements.
    \item {\sc SpenderQ} also provides latent variable encoding, or compression, 
    of the quasar spectra that are redshift invariant and physically informative.
    This encoding can be used to identify interesting anomalies 
    in the data. 
    As a demonstration, we show that the encodings can be used to
    classify quasars with BAL.
\end{enumerate}

In this work, we assess and benchmark {\sc SpenderQ} on mock spectra where 
we know the true intrinsic quasar spectra. 
We focus on how {\sc SpenderQ} significantly improves on the current
state-of-the-art in recovering the quasar continuum.
In subsequent works, we will again use the mocks to examine and quantify how 
the {\sc SpenderQ} improvements propagate to reducing distortions in the 
\lya~forest clustering measurements and, then, the BAO and full-shape 
clustering cosmological analyses. 

Furthermore, in an accompanying paper we apply {\sc SpenderQ} to 
DESI quasar spectra in the public 
Early Data Release~\citep{desi_kp1_sv, desi_kp1_edr}.
We present and publicly release the catalogs of the \lya~forest, metal absorbers,
latent variables, and reconstructed quasar spectra all constructed using 
{\sc SpenderQ}. 
Furthermore, we present how {\sc SpenderQ} reconstructions can be 
used to study the physics of quasars.
In subsequent works, we will use the latent variable catalog to identify 
outliers among the DESI quasars and also use the reconstructed quasar spectra 
to examine the variability of quasar spectra.
We will also use the \lya~forest catalog for unbiased \lya~BAO and full-shape 
clustering cosmological analyses.

%% file: supp.tex
\begin{figure}
  \centering
  \includegraphics[width=0.5\linewidth]{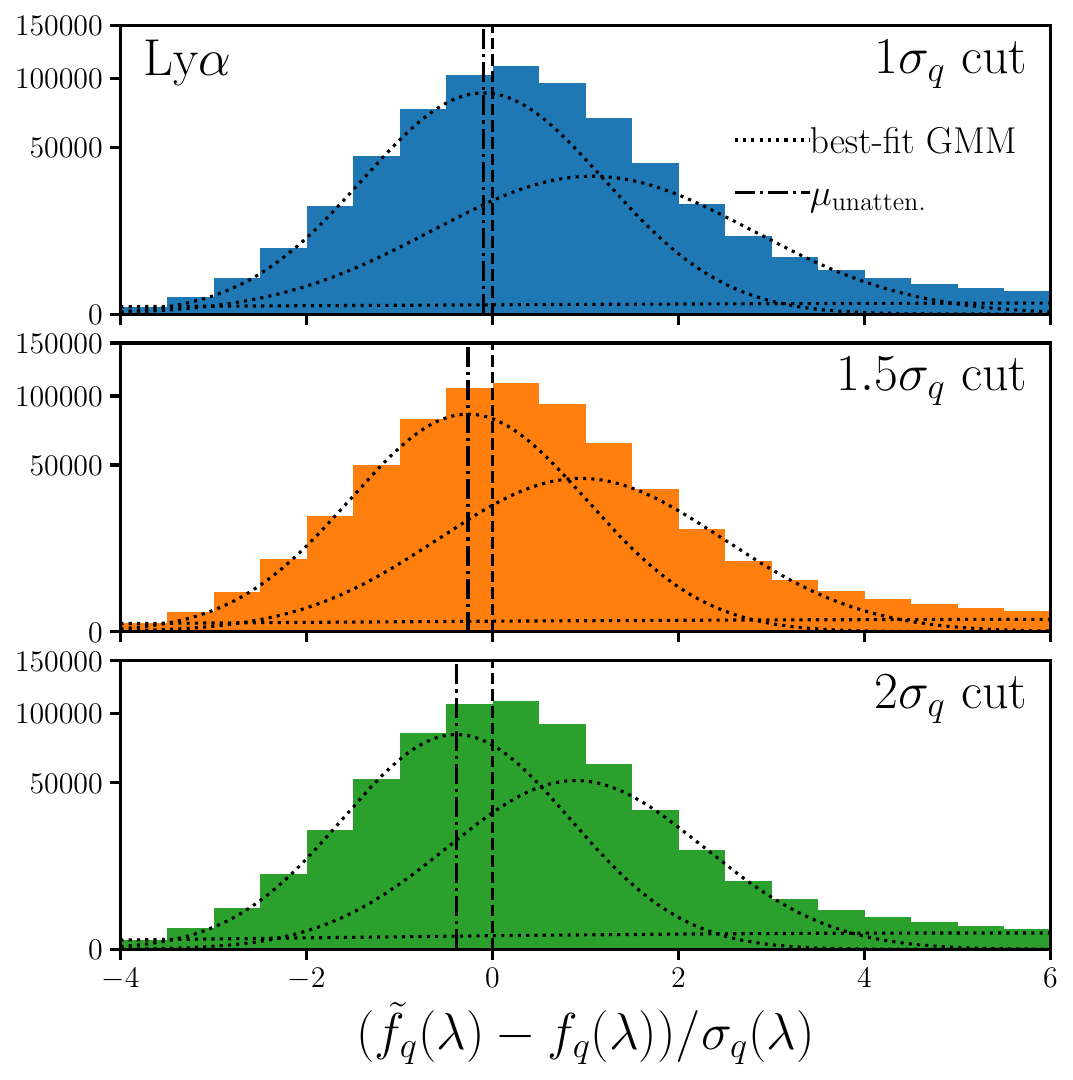}
  \caption{
  The distribution of residuals between the {\sc SpenderQ} reconstructions, $\tilde{f}_q$,
  and the observed spectra, $f_q$, normalized by observed uncertainty, $\sigma_q$. 
  We mark the components of the best-fit GMM (dotted) and the mean of
  the component that corresponds to the unattenuated continuum 
  ($\mu_{\rm unatten}$; dot-dashed).
  We also include $(\tilde{f}_q - f_q)/\sigma_q = 0$ (dashed) for reference.
  The y-axis is scaled by the square function to clearly show all three 
  GMM components. 
  We use these distributions to determine the threshold used to identify 
  the spectral elements 
  } \label{fig:sigma}
\end{figure}

\section{{\sc SpenderQ} Absorption Identification} \label{app:thresh}
One key part of the {\sc SpenderQ} framework is using the {\sc Spender} quasar continuum
reconstructions, $\tilde{f}_q$, to identify and mask parts of the spectra with absorption
for the next iteration. 
To do this, we compare $\tilde{f}_q$ to $f_q$ and identify the spectral elements where 
$f_q$ lies below $\tilde{f}_q$ by more than some threshold that depends on 
$\sigma_q(\lambda)$, the observed uncertainty: $f_q < \tilde{f}_q - c \sigma_q $.
In principle, the threshold, $c$, should be set by the amount of absorption, with different 
values for different wavelength ranges (e.g. regions A and B). 
While $c$ can be calibrated on simulations, we opt for a data-driven approach to make
{\sc SpenderQ} independent from assumptions that go into simulations. 

To determine the threshold, we first run {\sc SpenderQ} with different threshold 
values over a wide range, $c = 0.5 - 3$, until convergence. 
We then examine the distribution of $(\tilde{f}_q - f_q)/\sigma_q$, for the 
\lyb, \lya, and $\lambda > 1215$\r{A} wavelength ranges.
Next, we fit Gaussian mixture models~\citep[GMM;][]{mclachlan2000} to the distributions.
Then we identify the Gaussian component that correspond to the unattenuated continuum 
and select the $c$ value where the unattenuated continuum component is centered
most closely at 0. 

In Figure~\ref{fig:sigma}, we show the distributions of $(\tilde{f}_q - f_q)/\sigma_q$ for 
the \lya~wavelength range for different thresholds: $c = 1$ (top), 1.5 (center), and 2 (bottom). 
The distributions have a consistent shape that is well-described by three 
Gaussian components.
We mark the Gaussian components for the distributions in each panel (black dotted)
and include $(\tilde{f}_q - f_q)/\sigma_q = 0$ (dashed) for reference.
One component, left-most, corresponds to the unattenuated continuum with $\sigma\approx1$. 
The other two describe the attenuated spectral elements.
We scale the y-axis by the square function to clearly show all 
three GMM components.

The mean corresponding to the unattenuated continuum component 
($\mu_{\rm unatten.}$; black dot-dashed)
vary for the different thresholds. 
For the \lya~range, a threshold of $\sim1\sigma_q$ results in the unattenuated continuum 
component to be centered at 0.
This means that with this threshold, we accurately reconstruct the amplitude of the overall 
quasar continuum.
We can use this approach because {\sc SpenderQ} reconstructions capture 
the intrinsic shape of the quasar continuum for individual spectra. 
This is demonstrated by the consistent shapes of the $(\tilde{f}_q - f_q)/\sigma_q$ 
distributions for different $c$. 

\section{{\sc SpenderQ} Limitations}\label{app:limit}
\begin{figure}
  \centering
  \includegraphics[width=0.8\linewidth]{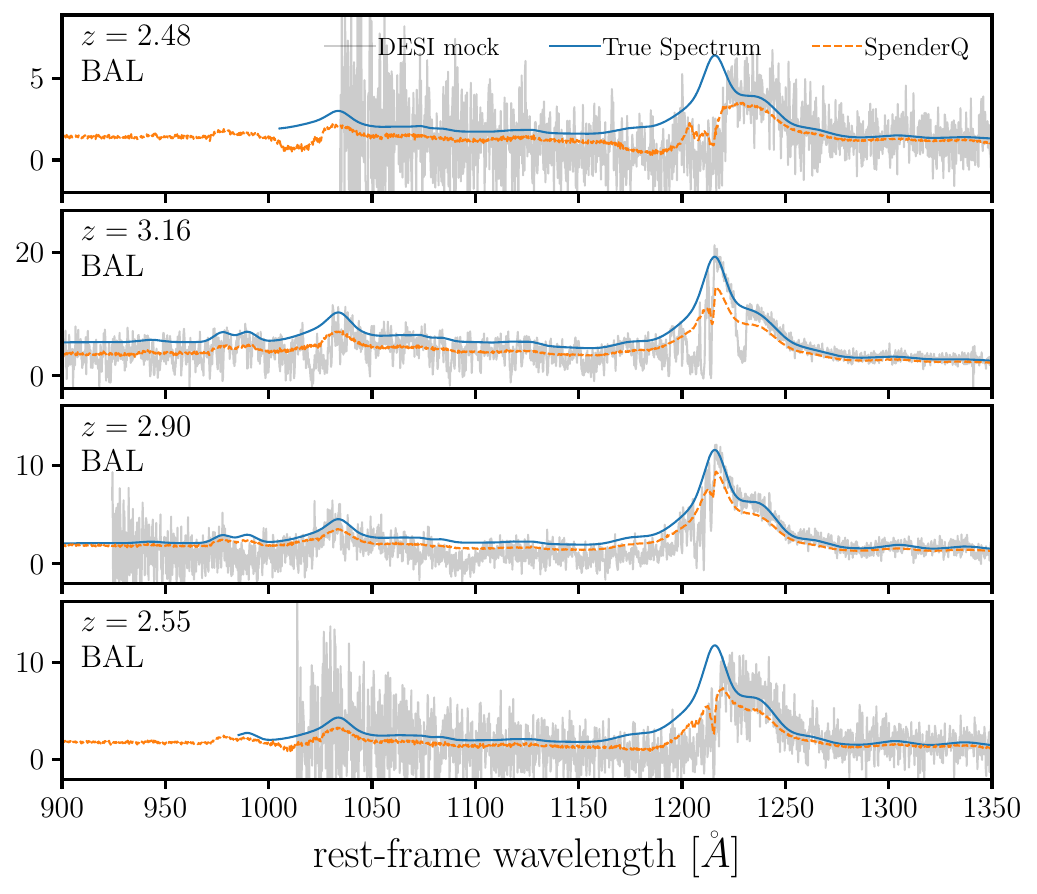}
  \caption{Higher signal-to-noise (SNR > 0.25) spectra where the median 
  {\sc SpenderQ} residual in the \lya~range exceeds 25\%.
  All of the spectra have broad absorption features in the 
  \lya~emission from BAL. 
  In the standard approach, spectra with these features are identified
  and removed in preprocessing steps before the reconstruction with 
  \picca. 
  In {\sc SpenderQ}, we can either use the same preprocessing steps or 
  these spectra can be identified in the latent space, as we show
  in Figure~\ref{fig:latents}.
  } \label{fig:bad_recon}
\end{figure}

Overall, {\sc SpenderQ} provides accurate reconstruction of individual quasar
continua. 
However, for a small fraction of the spectra (0.25\%) the median {\sc SpenderQ}
fractional residual exceed 25\% in the \lya~range.
The vast majority (99.2\%) of these are spectra with median SNR per pixel < 1. 
The rest are spectra with higher SNR but with major broad absorption 
features in the \lya~emission from BALs. 
In Figure~\ref{fig:bad_recon}, we present examples of these spectra, 
i.e., four spectra with higher SNR > 0.25.
All of the spectra have BALs in their \lya~emission.

\begin{figure}
  \centering
  \includegraphics[width=0.5\linewidth]{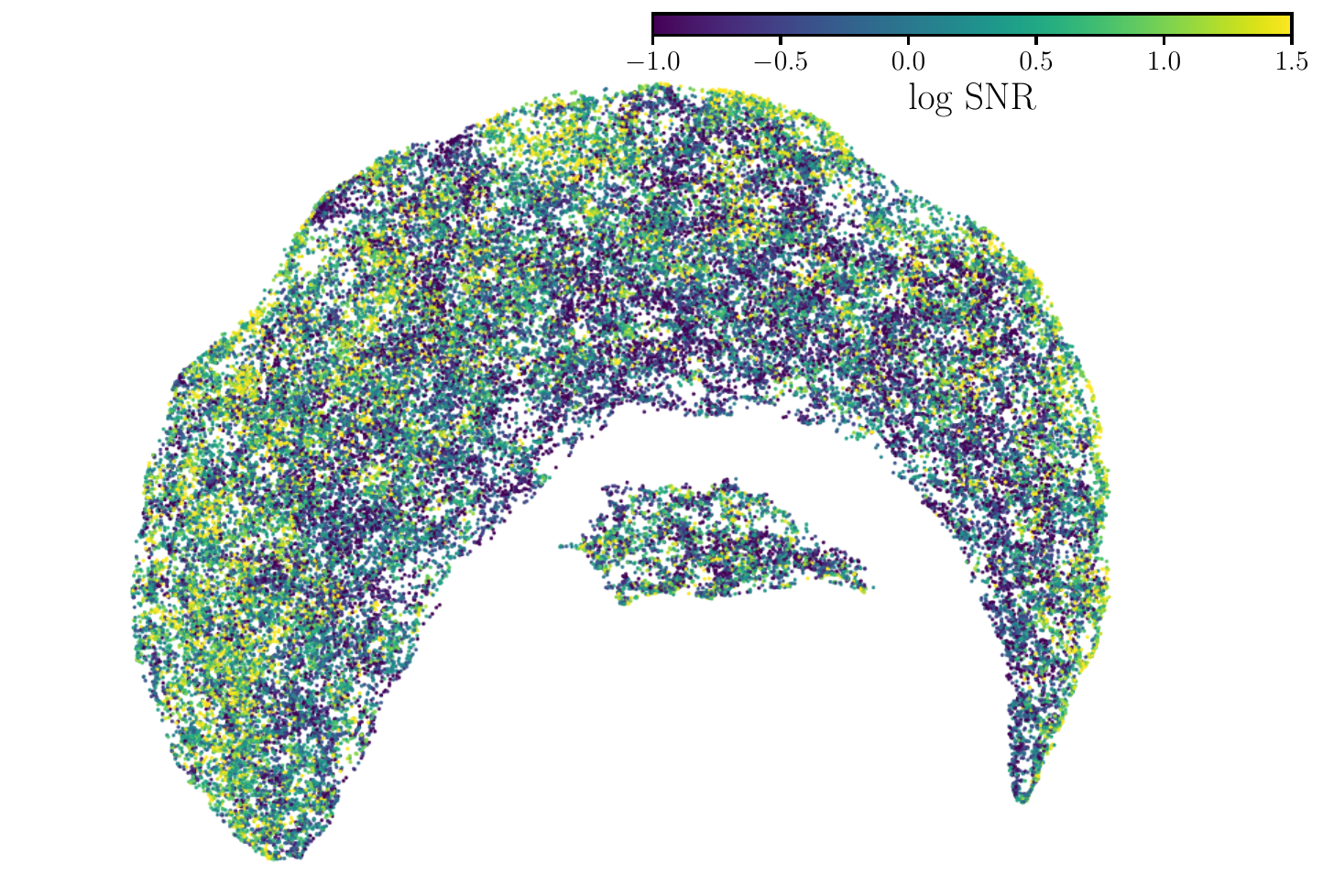}
  \caption{Two-dimensional UMAP embedding of the {\sc SpenderQ} 
  latent variables color coded by SNR. 
  The SNR represent the median SNR per spectral element over the 
  full wavelength range.
  Overall, we find no significant dependence of the latent space with
  SNR.} \label{fig:umap_snr}
\end{figure}

We further examine whether the SNR of the spectra impacts the
latent encoding of {\sc SpenderQ}.
In Figure~\ref{fig:umap_snr}, we present the two-dimensional UMAP 
embedding of the {\sc SpenderQ} latent variables, color-coded by 
the median SNR per pixel of the full spectra.
Overall, we find little SNR dependence in the latent space.
Interestingly, if we examine the SNR dependence together with the 
redshift dependence in the left panel of Figure~\ref{fig:latents}, 
we find that the lower portion of the upper cluster is 
occupied by lower redshift quasars with lower SNR spectra. 
This suggests that there are fainter quasars at lower redshift
that are not in the sample at higher redshifts. 
The {\sc SpenderQ} latent space suggests that they are intrinsically 
distinct from the higher redshift quasars in the sample.

